\documentclass[sigconf]{cidr-2027}

\usepackage{amsmath}
\usepackage{algorithmic}
\usepackage{caption}
\usepackage{graphicx}
\usepackage{textcomp}
\usepackage{xcolor}
\usepackage{multirow}
\usepackage{lipsum}
\usepackage{circledsteps}
\usepackage{threeparttable}
\usepackage{subfigure}
\PassOptionsToPackage{capitalize,noabbrev,nameinlink}{cleveref}
\usepackage{cleveref}
\usepackage{tabularx}
\usepackage{color}
\usepackage{colortbl}
\usepackage{balance}
\usepackage{wrapfig}
\usepackage{tcolorbox}
\usepackage{listings}
\usepackage{float}
\newfloat{lstfloat}{htbp}{lop}
\floatname{lstfloat}{Listing}
\crefformat{lstfloat}{#2Listing~#1#3}
\Crefformat{lstfloat}{#2Listing~#1#3}
\usepackage{mdframed}
\mdfdefinestyle{listing}{%
 hidealllines=true,leftline=true,
 innertopmargin=-1ex,%
 innerbottommargin=-1ex,%
 innerrightmargin=0em,%
 rightmargin=0em,%
 innerleftmargin=-1ex,%
 leftmargin=0ex,%
 middlelinewidth=.2em,%
 linewidth=2pt,
 linecolor=gray,
}
\usepackage{tikz}
\definecolor{tumblue}{rgb}{0,0.396,0.7412}
\newcommand*\rounded[1]{\tikz[baseline=(char.base)]{
      \node[shape=circle,inner sep=1pt,fill=tumblue] (char) {\textcolor{white}{#1}}}}
\usepackage{threeparttable}
\usepackage{color}
\usepackage{colortbl}
\usepackage{cleveref}
\usepackage{tabularx}
\usepackage{lipsum}
\usepackage{makecell}
\usepackage{enumitem}
\usetikzlibrary{shapes.arrows}
\newcommand{\whitearrow}{%
    \begin{tikzpicture}[baseline=(char.base)]
        \node(char) at (0,0) {\phantom{X}}; 
        \node [single arrow, draw, inner sep=1.5pt,
               minimum height=0.7em, 
               single arrow head extend=2.5pt,
               fill=white,
               line width=0.2pt, 
               anchor=center] at (char.center) {};
    \end{tikzpicture}%
}

\newcommand{\circhtag}[1]{%
  \tikz[baseline=(char.base)]{
    \node[draw,circle,inner sep=1pt] (char) {\footnotesize\textbf{\##1}};
  }%
}
\begin{document}

\title{The Case for Single-Version Concurrency Control in the Cloud}

\author{Lam-Duy Nguyen}
\email{lamduy.nguyen@tum.de}
\affiliation{%
  \institution{Technische Universität München}
  \country{Germany}
}
\author{Gabriel Haas}
\email{gabriel.haas@tum.de}
\affiliation{%
  \institution{Technische Universität München}
  \country{Germany}
}
\author{Muhammad El-Hindi}
\email{muhammad.el-hindi@tum.de}
\affiliation{%
  \institution{Technische Universität München}
  \country{Germany}
}
\author{Viktor Leis}
\email{leis@in.tum.de}
\affiliation{%
  \institution{Technische Universität München}
  \country{Germany}
}
\author{Philip A. Bernstein}
\email{phil.bernstein@microsoft.com}
\affiliation{%
  \institution{Microsoft Research}
  \country{USA}
}

\renewcommand{\shortauthors}{Lam-Duy Nguyen et al.}

\def\para#1{\noindent\textbf{#1.}}
\def\questionpara#1{\noindent\textbf{#1?}}


\begin{abstract}
Multi-version concurrency control (MVCC) has been the dominant concurrency protocol in modern database management systems (DBMSs).
However, cloud-native HTAP architectures shift many of MVCC's traditional advantages from the primary transaction engine to the deployment architecture.
Now, transactional and analytical processing are decoupled into loosely synchronized engines, while cloud deployment primitives provide the workload isolation required for this design.
In this paper, we challenge the status quo, arguing that single-version concurrency control (SVCC) is better suited to cloud transaction processing than MVCC.
Through a systematic analysis of MVCC's limitations, we show that SVCC avoids these shortcomings through a more lightweight and efficient design.
Beyond arguing for a lighter transaction processing, we use this SVCC effort as a blueprint to advocate for a broader rethinking of cloud-native architectures and to outline a research roadmap to shape next-generation DBMSs.
\end{abstract}





\makeatletter
\newcount\@authorbx@n
\@authorbx@n=0

\maketitle

\section{Introduction}
\label{sec:intro}

\para{MVCC has been the standard}
As of 2026, the five top-ranked database management systems (DBMSs) in the \href{https://db-engines.com/en/ranking}{DB-Engines ranking} -- Oracle, MySQL, SQL Server, PostgreSQL, and MongoDB -- all implement multi-version concurrency control (MVCC).
The appeal of MVCC over single-version concurrency control (SVCC) is that readers never block writers, so one engine can serve hybrid transactional and analytical processing (HTAP) workloads without a separate analytical system.
To gain that capability, DBMSs accept the substantial complexity and space overhead of maintaining multiple versions.

\begin{figure}[t]
\centerline{\includegraphics[width=\linewidth]{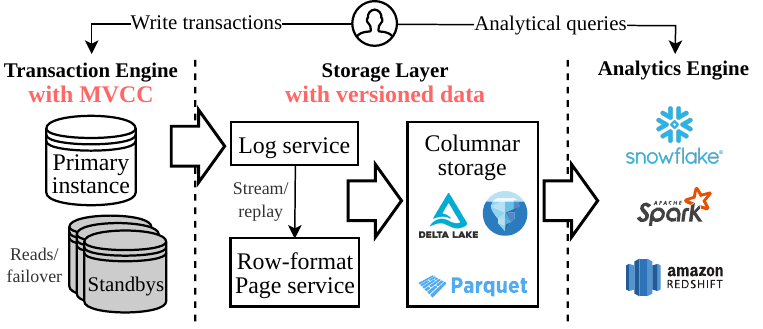}}
  \caption{Major cloud data platforms process transactions in a single-writer transaction engine (left) and queries in a separate analytics engine (right), with updates flowing through the storage layer (middle).
  Versioning in both the transaction engine and storage layer creates unnecessary overhead.
  \protect\whitearrow\; shows the primary data path.}
  \label{fig:cloud-db}
\end{figure}

\para{The cloud changes design assumptions}
The advent of cloud fundamentally changes the design objectives of database systems.
Cloud platforms offer elastic compute resources and fast instance provisioning, enabling multiple database systems to be deployed and tailored to different workloads~\cite{DBLP:journals/pvldb/LeisK21, DBLP:conf/icde/BonczCFHL0NSSZ23, DBLP:journals/pvldb/ZhangZM25, DBLP:conf/cidr/ZhangLY24}.
These properties naturally encourage a separation between transactional and analytical deployments.
The transaction side typically follows the single-writer architecture.
In this design, a primary instance (\cref{fig:cloud-db}, left) handles write transactions, e.g., PostgreSQL, while analytical reads are served in a dedicated analytical engine (\cref{fig:cloud-db}, right)~\cite{DBLP:journals/vldb/HaubenschildL25, DBLP:conf/sigmod/VerbitskiGSBGMK17, ltap, DBLP:conf/sigmod/AntonopoulosBDS19, DBLP:conf/sigmod/DepoutovitchCCL20, DBLP:journals/pvldb/MilkaiYP25, DBLP:journals/pvldb/GubnerHL25, DBLP:journals/tkde/ZhangLZZF24}.
This deployment is supported by row-oriented log and page services, with columnar storage populated through decoupled ETL pipelines or asynchronous transactional log streams, as shown in~\cref{fig:cloud-db} (middle).
Making this deployment model practical at scale requires efficient replication between components; recent advances in log-based replication have addressed this challenge, significantly reducing latency and overhead~\cite{DBLP:journals/pvldb/MilkaiYP25, DBLP:journals/pvldb/GubnerHL25, BtrLog}.
Altogether, these developments enable workload separation at the deployment level, reducing the need for complex MVCC mechanisms to support long-running snapshots and HTAP within the primary instance.

\para{MVCC adds overhead in single-writer cloud OLTP}
Given the single-writer model in cloud OLTP systems, let us now reassess the role of MVCC.
Under MVCC, each update creates a new physical version of a logical tuple, increasing memory pressure, storage footprint, and log volume.
Many mechanisms that add CPU overhead, reduce cache efficiency, and performance instability are consequences of this design: garbage collection, version visibility checks, and version chain traversal~\cite{DBLP:conf/sigmod/KimKCYKJ21, DBLP:journals/pvldb/AlhomssiL23}.
While these costs are justified in traditional systems that rely on MVCC for concurrent reads and writes, they are less important in single-writer cloud architectures.
The cloud already provides workload isolation at the deployment level through the storage layer or replicas, i.e., it conceptually resembles multi-versioning, making much of MVCC's benefits and associated overheads unnecessary on the primary instance.

\para{SVCC is a better fit for cloud OLTP}
In contrast, SVCC aligns naturally with the single-writer model.
By keeping a single physical version per object, it reduces memory and storage overhead and simplifies the primary engine.
As shown in~\cref{fig:cloud-db}, workload separation is handled at the deployment level: analytics run on a separate engine, while the primary focuses on transactions.
Additionally, if the primary instance can deliver high-performance serializability, the need to support weaker isolation levels inside the core engine for performance reasons vanishes, rendering one of MVCC's primary historical advantages unnecessary.
Performance stability also improves because SVCC eliminates MVCC's garbage collection and version traversal, avoiding the overhead those mechanisms incur.
As a result, the primary OLTP engine remains lightweight; most of the system complexity is now shifted to the deployment model, which is naturally enabled by cloud primitives.

\para{Contributions}
This paper re-examines concurrency control in the cloud era, focusing on the protocol used by the primary writer in a single-writer design.
Specifically, we challenge the conventional wisdom that MVCC should be the default scheme for the primary writer, and instead argue that single-versioning is better suited for it.
Our contributions are:

\begin{itemize}[leftmargin=*]
\item \textbf{\cref{sec:background} -- Revisiting MVCC:}
We briefly discuss the historical background that led to the widespread adoption of MVCC.
We then analyze MVCC in detail and expose the overheads MVCC incurs to achieve its advantages.
\item \textbf{\cref{sec:cloud} -- Cloud-centric analysis and SVCC positioning:}
We analyze how cloud characteristics reshape concurrency control trade-offs.
We then argue in detail that SVCC more closely aligns with cloud-native transaction processing than MVCC, offering lower overheads and reduced system complexity for OLTP.
\item \textbf{\cref{sec:future_work} and \cref{sec:related_work} -- Research directions:}
We discuss related work and outline possible research opportunities to improve cloud-native HTAP architectures beyond the SVCC focus.
\end{itemize}

\section{MVCC: The Good and the Ugly}
\label{sec:background}

The importance of managing concurrent access in DBMSs was recognized early~\cite{DBLP:journals/cacm/Codd70}.
The key to that is the transaction abstraction, proposed by Jim Gray et. al.~\cite{DBLP:conf/vldb/GrayLPT75}.
Jim Gray published many more papers on transaction processing~\cite{DBLP:conf/vldb/Gray81, DBLP:books/mk/GrayR93, DBLP:conf/icalp/Gray80, DBLP:conf/ds/GrayLPT76, berenson1995critique}, which emphasize the importance and benefits of transactions, and their atomicity, isolation (i.e., serializability) and durability (ACID) properties~\cite{DBLP:conf/sigmod/Bernstein25, DBLP:journals/corr/abs-2605-20466, DBLP:series/synthesis/2019Graefe}.

\subsection{The Good}

\para{The early prevalence of single-version design}
Many DBMSs soon followed and started integrating the transaction abstraction into their system.
In many early implementations, a transaction overwrote the tuple value directly~\cite{DBLP:journals/tods/StonebrakerWKH76, DBLP:conf/afips/HeldS75}, i.e., the system maintained only a single physical version of each record at any point in time.
This single-version design was both natural and practical.
Maintaining exactly one up-to-date copy of each tuple simplified storage management, recovery, and buffer management, making the approach straightforward to implement.
Moreover, it aligned with the hardware constraints of the time -- limited main memory and expensive magnetic disks -- where minimizing redundancy was essential.
Consequently, SVCC was the practical and economically sensible design choice in early database systems.

\para{The rise of multi-version storage}
In the mid-1980s, the price of magnetic disks declined to a point where data replication became economically feasible.
This shift in storage economics fundamentally changed the design space of database systems.
With storage capacity becoming less restrictive, DBMSs could now afford to maintain multiple physical versions of the same logical record~\cite{bernstein2023transactionpapersjimgray}.
Multi-version concurrency control preserves older versions of tuples instead of overwriting them in place~\cite{DBLP:conf/vldb/BernsteinG80a, DBLP:journals/tods/BernsteinG83, DBLP:phd/ndltd/Reed78}.
Consequently, long-running read queries can access older tuple versions and are not blocked by concurrent write transactions, significantly improving read-write concurrency and overall system performance.
Moreover, empirical studies show that many applications tolerate isolation levels weaker than full serializability~\cite{DBLP:conf/cidr/ChengHXBB023, DBLP:conf/sigmod/Pavlo17}.
MVCC naturally supports such relaxed isolation models (e.g., snapshot isolation) while reducing contention between readers and writers.
These practical advantages, combined with improved storage affordability, led to increased attention from database practitioners, which explains why MVCC was adopted by most modern DBMSs.

\para{Primary advantage: Non-blocking reads}
The main benefit of MVCC lies in its ability to decouple readers from writers.
By preserving multiple physical versions of a tuple and associating them with visibility metadata (e.g., version timestamps), long-running queries can access a consistent snapshot of the database without conflicting with concurrent update transactions.
Another commonly cited advantage of MVCC is \textit{time-travel queries}, which allow users to access historical database states; however, this feature introduces substantial performance and storage overhead, leading PostgreSQL to deprecate and eventually remove its time-travel support in v6.2~\cite{timetraveldisable}.
Such snapshot-based reads offer fresher views of the database than offloading reads to replicas in cloud-native DBMS architectures, which typically incur millisecond-scale replication lag~\cite{DBLP:conf/sigmod/VerbitskiGSBGMK17, DBLP:conf/sigmod/DepoutovitchCCL20, DBLP:conf/sigmod/AntonopoulosBDS19}.
In workloads where reads dominate, this leads to substantial improvements in throughput and tail latency stability.

\subsection{The Ugly}
\label{sec:mvcc_tradeoff}

With the prevalence of MVCC, why do we argue for SVCC in cloud transaction processing?
To answer, let us now discuss the trade-offs database designers must accept in exchange for MVCC's benefits.

\begin{figure}[t]
  \centerline{\includegraphics[width=\linewidth]{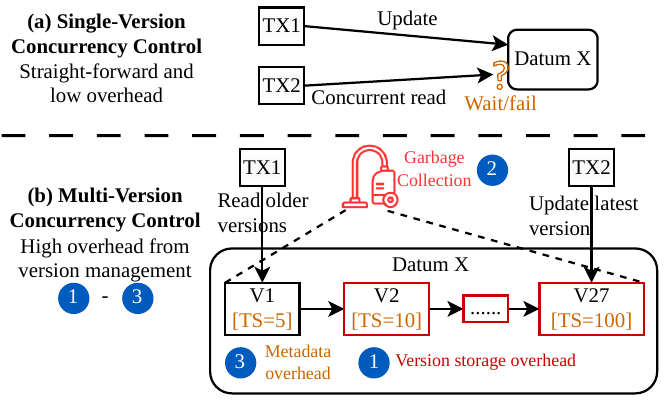}}
  \caption{Unlike SVCC, MVCC incurs additional overheads and complexity due to version management.}
  \label{fig:comparison}
\end{figure}

\para{Main trade-off: Complicated version management}
The fundamental cost of MVCC is that maintaining multiple versions per tuple requires additional machinery beyond the concurrency control algorithm (e.g., two-phase locking).
Unlike SVCC, MVCC treats every update as a version-creation event.
This fundamentally changes the responsibilities of the storage engine: Instead of managing a single mutable state per key, the system must continuously create, track, validate, and eventually reclaim multiple coexisting versions.
As a result, concurrency control is no longer isolated within the primary protocol.
It becomes tightly integrated with storage layout, indexing data structures, and memory management.
The complexity introduced by multi-versioning is therefore structural rather than incidental, as we illustrate in \Cref{fig:comparison} and examine in the following paragraphs.

\para{Overhead~\rounded{1}: Version storage}
Once updates produce new physical versions rather than overwriting in place, the system must decide where and how to store these versions for subsequent access.
Common approaches include undo-log-based version chains, append-only delta stores, or separate version indexes~\cite{DBLP:conf/sigmod/0001MK15, DBLP:journals/pvldb/WuALXP17, DBLP:journals/pvldb/AlhomssiL23, DBLP:conf/sigmod/KimKCYKJ21}.
While these designs differ in locality and write amplification, none avoids the fundamental cost of retaining multiple versions.

\para{Overhead~\rounded{1}~a: Version storage complexity}
Specifically, version storage introduces unavoidable indirection.
A lookup no longer focuses on only one record; instead, the engine must traverse the associated version chain to find a visible version.
Under write-heavy workloads, these chains can grow, degrading cache locality and increasing pointer chasing costs.
\citet{DBLP:journals/pvldb/AlhomssiL23} shows that preventing unbounded version accumulation in MVCC requires a combination of carefully engineered mechanisms, including (1) post-commit visibility validation to detect isolation conflicts, (2) auxiliary data structures for managing old versions to enable efficient sweeping and also shorten version chains, and (3) in-row storage of recent versions to improve performance; all of these are non-trivial, significantly increasing system complexity.
Studies in~\cite{DBLP:conf/sigmod/KimKCYKJ21, DBLP:journals/pvldb/WuALXP17} further demonstrate that preventing excessive MVCC version-chain growth is challenging and requires substantial modifications to the core execution engine.

\para{Overhead~\rounded{1}~b: Version storage for large objects}
Handling non-inline attributes (e.g., large VARCHAR, TEXT, JSON, or BLOB columns) further complicates version storage:
The value of such attributes is typically not stored inline with the tuple; instead, the tuple contains only a pointer to separately allocated storage, introducing an additional level of indirection to access the actual object~\cite{nguyen2024blob, DBLP:journals/pvldb/WuALXP17}.
Under MVCC, the DBMS may allow multiple versions to share the same external object; hence, reference counters are necessary to prevent excessive space consumption and ensure that values are reclaimed only when no version references them.
This significantly complicates \textit{garbage collection}, an already challenging component of MVCC as we will discuss next.

\para{Overhead~\rounded{2}: Garbage collection}
Because old versions may remain visible to active transactions, they cannot be reclaimed immediately.
MVCC, therefore, requires a garbage collection mechanism to determine when a version becomes globally unreachable and can be safely removed.
Garbage collection introduces several systematic trade-offs.
First, determining global visibility information requires coordination, introducing additional synchronization overhead~\cite{DBLP:journals/pvldb/AlhomssiL23, ScalableGC}.
Second, it requires background work that competes with transactions for CPU cycles, cache capacity, and I/O bandwidth.
Third, garbage reclamation policies are inherently sensitive to workload characteristics:
A single long-running transaction can delay cleanup indefinitely, leading to version accumulation and wasted space~\cite{DBLP:journals/pvldb/AlhomssiL23}.
Conversely, aggressive reclamation increases bookkeeping overhead and interferes with transaction processing.
Consequently, it is difficult to predict when garbage collection will occur and how much overhead it will incur, making it a persistent source of performance instability~\cite{ScalableGC}.
More importantly, garbage collection is not merely an optimization but a correctness requirement in MVCC.
Without it, version chains grow unbounded, index entries accumulate, and storage consumption increases significantly over time.
Thus, multi-versioning introduces a permanent maintenance responsibility that does not exist in SVCC.

\para{Overhead~\rounded{3}: Tuple metadata and secondary index}
Beyond version storage and garbage collection, MVCC imposes persistent space and runtime overheads due to tuple metadata and secondary indexes.
Each tuple must store visibility metadata (e.g., timestamps and/or transaction IDs), leading to a typical header overhead of 16-32 bytes or more per record~\cite{DBLP:journals/pvldb/AlhomssiL23, DBLP:journals/pvldb/WuALXP17}.
For narrow schemas, this metadata can constitute a substantial fraction of the tuple size, reducing effective storage capacity and cache efficiency.
Secondary index management is also more involved under MVCC, since both index keys and tuple pointers may change across versions.
Different design choices exhibit different performance trade-offs across workloads, and index-only scans are generally infeasible unless visibility metadata is embedded directly into index entries, further increasing space overhead~\cite{DBLP:journals/pvldb/WuALXP17}.

\section{SVCC: The Better Fit for the Cloud}
\label{sec:cloud}

Modern cloud data platforms comprise separate transactional and analytical engines connected through a disaggregated storage layer.
In this section, we first identify the design principles that underpin these platforms and reshape concurrency-control requirements.
We then show that these requirements render MVCC's primary overheads (c.f., \cref{sec:mvcc_tradeoff}) difficult to justify for the primary OLTP engine, making SVCC the more natural choice for cloud OLTP.

\subsection{New Design Principles in Cloud DBMSs}
\label{sec:design_principles}
Modern cloud data platforms follow three design principles that shape their high-level architecture and, in turn, their concurrency-control requirements.

\para{Cost and resource efficiency as a necessity}
In traditional on-premises deployments, databases ran on dedicated physical hardware, with the main goal being to maximize performance within fixed resource constraints.
In contrast, cloud computing follows a pay-as-you-go model, where CPU, memory, storage, and network resources are billed based on usage.
As a result, cost efficiency becomes a primary objective~\cite{DBLP:journals/pvldb/TanGPYSDSAK19, DBLP:journals/pvldb/MisegiannisRLG25, DBLP:conf/cidr/DuweALLL0T25, DBLP:conf/cidr/SteinertKL26}; any mechanism that consumes CPU time, memory, or storage must justify its cost.
Consequently, designs that eliminate unnecessary overhead while maintaining high performance become increasingly attractive.

\para{Deployment-level workload isolation}
The elasticity of the cloud also enables new architectural patterns.
On-demand instance provisioning~\cite{DBLP:conf/cidr/SteinertKL26, DBLP:conf/nsdi/VuppalapatiMATM20} and low-latency log-based replication~\cite{DBLP:journals/pvldb/MilkaiYP25, DBLP:journals/pvldb/GubnerHL25, DBLP:journals/vldb/HaubenschildL25} make it easy to run multiple instances of the same system.
Instead of serving all workloads within a single engine, operators can deploy a dedicated instance for transaction processing and offload long-running analytical queries to replica instances or a separate engine.
Workload isolation is thus provided at the deployment level rather than by the concurrency-control mechanism within an engine.

\para{The single-writer architecture}
In line with the principle of workload isolation, several major cloud DBMS vendors have adopted a \textit{single-writer} architecture in their cloud OLTP systems, with examples including Amazon Aurora~\cite{DBLP:conf/sigmod/VerbitskiGSBGMK17}, Microsoft Socrates~\cite{DBLP:conf/sigmod/AntonopoulosBDS19}, Huawei Taurus~\cite{DBLP:conf/sigmod/DepoutovitchCCL20}, and Alibaba PolarDB~\cite{DBLP:journals/pacmmod/WangLSYZLYWLYWC23}.
In this architecture, one primary instance handles all writes while multiple replicas act as failover~\cite{DBLP:journals/tkde/ZhangLZZF24, ltap}.
The primary writer can therefore be optimized for short OLTP transactions, while read-intensive and analytical workloads execute elsewhere.
The single-writer architecture thus reinforces the broader shift from engine-level to deployment-level workload isolation.

\para{Implications for concurrency control}
As discussed in~\cref{sec:background}, MVCC was designed to isolate different workloads within a single engine.
These MVCC features were necessary when one database instance had to handle heterogeneous workloads.
In cloud-native DBMSs, however, workload isolation can be achieved through replicas and elastic instances, with replica freshness typically within milliseconds~\cite{DBLP:conf/sigmod/VerbitskiGSBGMK17, DBLP:conf/sigmod/DepoutovitchCCL20, DBLP:conf/sigmod/AntonopoulosBDS19}.
Consequently, the primary OLTP engine no longer needs to execute long-running queries or provide cross-workload isolation.
Its concurrency-control requirements are therefore substantially narrower.
The relevant question is no longer how to support heterogeneous workloads within one engine, but which concurrency-control scheme best serves a dedicated OLTP writer.
Given these architectural changes, is MVCC still the right concurrency control mechanism for the primary OLTP engine?

\subsection{Why SVCC Fits Cloud OLTP}

The cloud deployment model removes a principal motivation for MVCC in the primary OLTP engine: supporting heterogeneous workloads and long-running snapshots within a single engine.
Once these workloads are isolated in dedicated analytical engines, the primary writer can focus on transactions.
Under these narrower requirements, SVCC is not only viable but also a more natural foundation for cloud OLTP.

\para{SVCC is cheaper at every level}
As discussed in~\cref{sec:mvcc_tradeoff}, MVCC introduces substantial engine-level complexity, resulting in runtime overheads that are not present in single-version designs.
In particular, MVCC requires intricate mechanisms to manage multiple versions of both standard data types (e.g., numeric fields) and large objects.
In contrast, SVCC avoids this additional machinery and relies only on the underlying concurrency control scheme, such as 2PL.
Moreover, SVCC preserves a simple and intuitive model that remains consistent across all layers of the system.

\para{Modern engines make serializability practical}
To achieve good performance in the past, many applications deliberately accept weaker isolation levels -- which are naturally enabled by MVCC-based designs~\cite{berenson1995critique, DBLP:conf/sigmod/0001MK15} -- tolerating anomalies such as phantom reads or non-repeatable reads~\cite{DBLP:conf/cidr/ChengHXBB023, DBLP:conf/sigmod/Pavlo17}.
However, we argue that applications do not have to make that trade-off anymore.
State-of-the-art prototype engines can now achieve hundreds of thousands (or more) of transactions per second under strict serializability on the primary instance~\cite{DBLP:journals/vldb/HaubenschildL25, DBLP:conf/sigmod/0001MK15, DBLP:conf/osdi/EldeebXBCY23, DBLP:conf/cidr/EldeebBCY24, zinsmeister2026predictive}.
The throughput headroom is now large enough that most applications can adopt serializability by default, making SVCC the natural choice without sacrificing the correctness guarantees DBMSs previously had to compromise.

\subsection{Complexity Evaluation}
\label{sec:maintenance}

\para{Setup}
In this section, we use our own implementation experience as a concrete proxy to represent complexity.
We use LeanStore~\cite{DBLP:journals/pvldb/Leis24}, a modern storage engine that supports high throughput with $\mu$s-scale commit latency~\cite{DBLP:journals/pacmmod/NguyenL25}, and implement both protocols:

\begin{itemize}[leftmargin=*]
    \item \textbf{SVCC}: We implement traditional two-phase locking (2PL) with Wait-Die deadlock prevention.
    We use an \href{https://github.com/rotaki/tpcc-runner/blob/master/protocols/waitdie/include/waitdielock.hpp}{open-sourced wait-die lock implementation} combined with Intel TBB concurrent hash tables as an in-memory wait-die lock manager.
    \item \textbf{MVCC}: We implement Hyper-style MVCC~\cite{DBLP:conf/sigmod/0001MK15}.
    In this protocol, a transaction modifies tuples in place, buffers their before-images in an undo log, and, at commit time, validates the writes of concurrently committed transactions against its read predicates.
\end{itemize}

\noindent In both implementations, we use key-level synchronization instead of range locking, e.g., predicate-based validation in Hyper~\cite{DBLP:conf/sigmod/0001MK15}, for simplicity of implementation.
Although this modification permits phantom reads, it does not affect our comparison, since our focus is on complexity rather than performance.

\begin{table}[t]
    \caption{Approximate complexity comparison.
    \textit{Main} measures implementation effort for core components, while \textit{Modif.} and \textit{Tests} capture integration and testing effort.
    }
    \begin{tabular}{lrrr}
        \toprule
        \textbf{Protocol} & \multicolumn{3}{c}{\textbf{Lines of code (LoC)}} \\
         \cmidrule(l{0pt}){2-4}
           & \textbf{Main} & \textbf{Modif.} & \textbf{Tests} \\
         \midrule

    	\multicolumn{4}{l}{\textit{Prototype implementations}} \\
    	\addlinespace[0.25em]
    	SVCC                                                    & 303       & 45  & 665  \\
    	MVCC                                                    & 426       & 265 & 1325 \\

    	\midrule
    	\multicolumn{4}{l}{\textit{Real implementations (AI-estimated)}} \\
    	\addlinespace[0.25em]
    	OSIC (Multi-VCC)~\cite{DBLP:journals/pvldb/AlhomssiL23} & $\sim$13K & --  & --   \\
    	PostgreSQL (Multi-VCC)~\cite{PostgreSQL}                & $\sim$35K & --  & --   \\
    	MyISAM (Single-VCC)~\cite{MySQLMyISAM}                  & $\sim$2K  & --  & --   \\
    	\bottomrule
    \end{tabular}
    \label{tab:complexity_cmp}
\end{table}

\para{Performance comparison}
The performance characteristics of both protocols, as well as many other concurrency control designs (i.e., which workloads they are optimized for) are well-studied elsewhere~\cite{DBLP:journals/pvldb/WuALXP17, DBLP:journals/vldb/BangMPB22, DBLP:conf/damon/BangMPB20, ye2023polaris, DBLP:conf/sigmod/ChenYKAAS22, guo2021releasing, DBLP:journals/pvldb/YuBPDS14, DBLP:journals/pvldb/TanabeHKT20}; we do not revisit them in depth here.
Our focus is on lightweight, clean prototypes that highlight the operational and implementation complexity of the two protocols.
We refer readers to the cited studies for comprehensive evaluations of those concurrency control protocols.

\para{Reduced complexity in the prototype}
\cref{tab:complexity_cmp} shows that SVCC comes with a consistently lower engineering effort across dimensions.
SVCC requires 303 LoC for core implementation and 45 LoC for modifications, while MVCC needs 426 and 265 LoC, respectively, mainly due to version-chain integration with indexes.
The gap is even larger in testing, where MVCC nearly doubles the code size (1325 vs.~665 LoC).
Overall, MVCC incurs about 2--3$\times$ higher engineering cost.
Furthermore, this difference is not just quantitative.
MVCC requires developers to reason simultaneously about version chains, garbage collection, and their interactions with other subsystems.
In contrast, SVCC provides a more uniform abstraction, simplifying both implementation and validation.

\para{Complexity comparison of real-world implementations}
Our prototype already suggests that SVCC requires substantially less implementation effort than MVCC (\cref{tab:complexity_cmp}).
To assess whether this trend extends to production systems, we compare a few representative implementations.
For MVCC, we consider Ordered Snapshot Instant Commit (OSIC)~\cite{DBLP:journals/pvldb/AlhomssiL23} and PostgreSQL~\cite{PostgreSQL}; for single-version storage, we consider MyISAM~\cite{MySQLMyISAM}.
Using ClaudeAI to estimate implementation complexity, we find that OSIC comprises approximately 13K LoC, PostgreSQL roughly 35K LoC for MVCC, whereas MyISAM requires only around 2K LoC.
While we acknowledge this is an apples-to-oranges comparison -- MyISAM uses table-level locking rather than row-level concurrency control -- we argue that the statistics still provide a useful reference point for highlighting the implementation complexity of MVCC.
Consistent with this observation, (1) a popular PostgreSQL internals book~\cite{rogov2023postgresql} devotes 108 pages (about 20\% of its content) to transactions and MVCC, and (2) practitioners have noted that MVCC leaves its signatures across nearly every subsystem of PostgreSQL~\cite{PostgreMVCCBlog1}.
Together with our analysis, these observations suggest that cloud-native DBMSs can benefit from the simpler SVCC, which achieves lower implementation complexity in our study.

\subsection{Lightweight Extensions}

\begin{figure}
    \begin{center}
        \includegraphics[width=0.95\columnwidth]{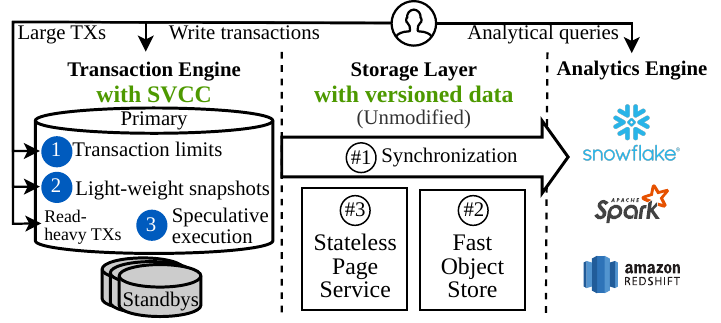}
    \end{center}
    \caption{Our target architecture uses SVCC in the primary instance. Lightweight mechanisms are used to handle large transactions or read-heavy queries (i.e.,  \protect\rounded{\#}).
    Future directions for cloud HTAP are outlined as~ \protect\circhtag{}.}
    \label{fig:target-architecture}
\end{figure}

Building on the cloud-native architecture described in~\cref{fig:cloud-db}, the primary instance now adopts SVCC as its concurrency control protocol (c.f., \cref{fig:target-architecture}).
The lightweight design of SVCC intentionally scopes the primary engine to short transactions, which is the typical assumption for OLTP workloads.
A transaction may, however, have a large write set (e.g., bulk loading), scan a large amount of data before performing an update, or remain active for an extended period.
While such long-running transactions are a known challenge within the concurrency control community~\cite{DBLP:conf/sigmod/KimCKYKJ20, DBLP:journals/pvldb/AlhomssiL23}, SVCC's simplicity leaves room for lightweight extensions to accommodate these exceptional cases.
Below, we discuss several such extensions that we leave for future work.

\para{\rounded{1} Transaction limits}
Long-running transactions are challenging under both single-version and multi-version storage.
Depending on the protocol, such transactions retain locks, old versions, and/or the dependency graph for longer than typical transactions, thereby throttling the whole system.
This also explains why recent works propose to lock the entire database for long-running transactions~\cite{DBLP:conf/cidr/NeumannF20, DBLP:journals/pvldb/NguyenLN22}.
Production systems, therefore, commonly bound transaction size and duration.
Amazon Aurora DSQL limits the data modified by a write transaction to $10\,\mathrm{MiB}$ and the duration of the transaction to five minutes.
FoundationDB limits affected data to $10\,\mathrm{MB}$ and transaction duration to five seconds~\cite{DBLP:journals/corr/abs-2607-13276}.
These restrictions reflect a pragmatic design choice: exceptional operations should be decomposed or handled separately rather than determining the common-case design of the transaction engine.

\para{\rounded{2} Lightweight snapshots}
Another direction is to use a two-version extension on top of SVCC (e.g., 2V2PL~\cite{bernstein1987concurrency}, 2VTO~\cite{DBLP:journals/pvldb/NguyenLN22}).
However, these approaches still carry some of the overhead associated with MVCC.
A cheaper alternative is to provide snapshots on demand:
A transaction calls \texttt{BeginSnapshot()} to establish an empty auxiliary index -- all overwritten tuples from concurrent transactions are inserted into this index, and \texttt{EndSnapshot()} discards this auxiliary index completely when that snapshot is no longer needed.

\para{\rounded{3} Speculative execution}
Some transactions cannot be classified as read-only in advance -- they may perform substantial computation before their first write.
One possible approach is to begin such a transaction speculatively as read-only and upgrade it to a write transaction upon its first write.
This upgrade operation needs to validate all earlier reads.
If validation fails, the transaction can restart; alternatively, the upgrade may involve locking the whole database during the transition, similar to prior works~\cite{DBLP:journals/pvldb/FreitagKN22, DBLP:journals/pvldb/NguyenLN22}.

\section{Beyond SVCC: Rethinking Cloud HTAP}
\label{sec:future_work}

Our experience with SVCC suggests that many of the complexities in modern cloud-native DBMSs arise from legacy design assumptions rooted in traditional DBMS architectures.
We therefore believe this is the time to step back and consider the broader system.
Through this looking glass, we believe that the community has the opportunity to identify a leaner and more efficient architectural blueprint for cloud-native DBMSs.
Many existing proposals~\cite{DBLP:journals/pvldb/ChenDLLZZWCZLZS22, zaharia2025bringing, DBLP:conf/sigmod/ProutWVSLCBHWGS22, DBLP:journals/pvldb/YangRXSYLZZMCGD20, DBLP:journals/pacmmod/WangLSYZLYWLYWC23, DBLP:journals/tkde/DongZLZ24, DBLP:journals/pvldb/GubnerHL25} preserve these legacy assumptions, resulting in unnecessary complexity and overhead across the stack.
Below, we outline several research directions that we believe are key to building the next generation of cloud-native DBMSs.

\para{Direction \circhtag{1}: HTAP synchronization}
In traditional on-premise DBMSs, analytical queries rely on MVCC snapshots to achieve near-real-time visibility, as discussed in~\cref{sec:background}.
In cloud-native HTAP architectures, however, analytical snapshots are maintained outside the transactional engine, i.e., in storage layer -- \cref{fig:target-architecture} (middle), making snapshot synchronization latency a primary optimization objective.
Recent approaches reduce this lag by exposing recent updates through row-oriented page services~\cite{ltapdoc} or log-caching layers~\cite{DBLP:journals/pvldb/MilkaiYP25}.
Both approaches face the same issue: when the transactional engine produces a large volume of updates between the last synchronization point and the analytical snapshot timestamp, the analytical engine must merge all accumulated logs before answering queries, potentially negating the benefits of engines such as DuckDB or Raydon if the log gap is large (e.g., GBs).
We therefore believe that efficient HTAP synchronization remains an open research problem.
One direction is partitioning the log (and page) service by analytical predicates, allowing analytical engines to retrieve only relevant partitions and reducing unnecessary data movement.

\para{Direction \circhtag{2}: SSD-based object stores}
Complementing software-level log partitioning, redesigning the underlying cloud object stores (e.g., Amazon S3) offers a hardware-aware path to reducing HTAP synchronization latency.
Contemporary cloud object stores are inadequate not only for cloud-native HTAP systems but also for database workloads more broadly.
This limitation stems from the fact that today's object stores largely inherit architectures designed nearly two decades ago and have not been fundamentally rearchitected to take advantage of modern hardware.
Recent developments, such as S3 Express, reflect the growing demand for high-performance object storage; however, these remain incremental enhancements to the existing object-store abstraction and still incur access latencies that are too high for database workloads.
We envision new SSD-based object-store primitives designed from the ground up for cloud-native DBMSs, preserving the scalability and durability properties while providing substantially lower access latency.

\para{Direction \circhtag{3}: Stateless page service}
Another direction for optimizing cloud-native HTAP architectures is to make page materialization truly elastic.
As analytical workloads become more intensive, the page service must scale elastically to sustain high request throughput.
This, in turn, requires page service instances to join and leave as fast as possible.
However, existing cloud page services (e.g., Databricks/Neon LTAP~\cite{ltapdoc}) maintain dirty pages and asynchronously flush them to object storage, requiring coordination across page service instances.
We propose a \textit{truly stateless} page service based on the \textit{log-is-the-data} paradigm.
Durability, replication, and persistence are handled entirely by the log service, while buffer evictions and checkpointing append pages directly to the log service.
As a result, page service instances become transient execution units that can be created, assigned to table predicates, and removed independently; this also enables page service deployment on serverless infrastructure.
Thus, page materialization becomes a fully elastic compute layer rather than a stateful storage.


\section{Related Work and Summary}
\label{sec:related_work}

\para{MVCC complexity}
The complexity of MVCC protocols is well recognized.
\citet{DBLP:journals/pvldb/WuALXP17} show that MVCC incurs substantial overhead and complexity from version maintenance, garbage collection, and visibility checks.
Complementary results in \cite{DBLP:journals/pvldb/TanabeHKT20} demonstrate that MVCC only outperforms SVCC in limited cases (e.g., high-contention workloads).
Overall, these findings suggest that MVCC’s benefits are neither universal nor free, while introducing significant complexity and overhead.

\para{Lock manager for storage-based DBMSs}
Existing designs face a fundamental trade-off: they either assume that datasets fit entirely in memory and therefore do not support out-of-memory workloads~\cite{DBLP:journals/tods/JungHFHY14, DBLP:journals/vldb/RenTA15}, or they scale poorly and/or incur excessive storage overhead~\cite{DBLP:journals/pvldb/PortsG12}.
As SVCC regains attention, a new design that is both scalable and space-efficient for out-of-memory workloads remains an important open challenge.
A promising starting point is to adapt the per-page mapping table introduced by Umbra~\cite{DBLP:journals/pvldb/FreitagKN22}.

\para{Summary}
Our work is a first step toward the broader vision of how rethinking system components can reduce complexity while improving overall robustness.
With this paper, we hope to inspire further research on re-architecting the cloud database stack and encourages a broader community effort to build a leaner, more robust foundation for the next generation of cloud databases.


\balance
\bibliographystyle{ACM-Reference-Format}
\bibliography{references}

\end{document}